\documentclass[journal]{IEEEtran}
\pdfoutput=1
\IEEEoverridecommandlockouts

\usepackage{acronym}
\usepackage{amsmath,amssymb}
\usepackage[english]{babel}
\usepackage{bbm}
\usepackage{bm}
\usepackage{cite}
\usepackage{enumerate}
\usepackage{graphicx}

\usepackage{mathtools}

\usepackage{lscape}
\usepackage{longtable}
\usepackage{siunitx}
\usepackage{xcolor}
\usepackage{algorithm}
\usepackage{algorithmic}
\usepackage{rotating}

\usepackage[hidelinks]{hyperref}
\usepackage{cleveref}

\definecolor{myblue}{RGB}{31,119,180}
\definecolor{myorange}{RGB}{255,127,14}
\definecolor{mygreen}{RGB}{44,160,44}

\acrodef{AHL}{$N$-acyl homoserine lactones}
\acrodef{AI-2}{autoinducer 2}
\acrodef{ATP}{adenosine triphosphate}
\acrodef{BER}{bit error rate}
\acrodef{bioFET}{field-effect transistor biosensor}
\acrodef{C6-HSL}{N-(3-Oxyhexanoyl)-L-homoserine lactone}
\acrodef{CIR}{channel impulse response}
\acrodef{CSK}{concentration shift keying}
\acrodef{DNA}{deoxyribonucleic acid}
\acrodef{EGFET}{electrolyte-gated field-effect transistor}
\acrodef{EM}{electromagnetic waves}
\acrodef{FRET}{F\"orster Resonance Energy Transfer}
\acrodef{FSK}{frequency shift keying}
\acrodef{GFP}{green fluorescent protein}
\acrodef{IoBNT}{Internet of Bio-Nano-Things}
\acrodef{ISI}{inter-symbol interference}
\acrodef{LoC}{lab-on-a-chip}
\acrodef{MC}{Molecular Communication}
\acrodef{MIMO}{multiple-input and multiple-output}
\acrodef{NMC}{Natural Molecular Communication}
\acrodef{OOK}{on-off keying}
\acrodef{RSK}{reaction shift keying}
\acrodef{RNA}{ribonucleic acid}
\acrodef{Rx}{receiver}
\acrodef{sfGFP}{super-folded green fluorescent protein}
\acrodef{SISO}{single-input and single-output}
\acrodef{SMC}{Synthetic Molecular Communication}
\acrodef{SNR}{signal-to-noise ratio}
\acrodef{SPION}{superparamagnetic iron oxide nanoparticle}
\acrodef{ssDNA}{single-stranded DNA}
\acrodef{Tx}{transmitter}
\acrodef{VOC}{volatile organic compound}

\allowdisplaybreaks

\begin{document}
        
\title{Experimental Research \\in Synthetic Molecular Communications -  \\
Part I: Overview and Short-Range Systems}

\author{\IEEEauthorblockN{Sebastian Lotter\IEEEauthorrefmark{1},
    Lukas Brand\IEEEauthorrefmark{1},
    Vahid Jamali\IEEEauthorrefmark{2},
    Maximilian Sch\"afer\IEEEauthorrefmark{1},
    Helene M.~Loos\IEEEauthorrefmark{1}\IEEEauthorrefmark{4},
    Harald Unterweger\IEEEauthorrefmark{3},\\
    Sandra Greiner\IEEEauthorrefmark{1},
    Jens Kirchner\IEEEauthorrefmark{1},
    Christoph Alexiou\IEEEauthorrefmark{3},
    Dietmar Drummer\IEEEauthorrefmark{1},
    Georg Fischer\IEEEauthorrefmark{1},
    Andrea Buettner\IEEEauthorrefmark{1}\IEEEauthorrefmark{4},\\
    and Robert Schober\IEEEauthorrefmark{1}\\} 
    \IEEEauthorblockA{\small\IEEEauthorrefmark{1}Friedrich-Alexander-Universit\"at Erlangen-N\"urnberg (FAU), Erlangen, Germany}\\
    \IEEEauthorblockA{\small\IEEEauthorrefmark{2}Technical University of Darmstadt, Darmstadt, Germany}\\
    \IEEEauthorblockA{\small\IEEEauthorrefmark{4}Fraunhofer Institute for Process Engineering and Packaging IVV, Freising, Germany}\\
    \IEEEauthorblockA{\small\IEEEauthorrefmark{3}Universit\"atsklinikum Erlangen, Section of Experimental Oncology and Nanomedicine (SEON), Erlangen, Germany}
    \vspace*{-1cm}
}

\maketitle

\begin{abstract}
Since its emergence from the communication engineering community around one and a half decades ago, the field of \ac{SMC} has experienced continued growth, both in the number of technical contributions from a vibrant community and in terms of research funding.
Throughout this process, the vision of \ac{SMC} as a novel, revolutionary communication paradigm has constantly evolved, driven by feedback from theoretical and experimental studies, respectively.
It is believed that especially the latter ones will be crucial for the transition of \ac{SMC} towards a higher technology readiness level in the near future.
In this spirit, we present here a comprehensive survey of experimental research in \ac{SMC}.
In particular, this survey focuses on highlighting the major drivers behind different lines of experimental research in terms of the respective envisioned applications.
This approach allows us to categorize existing works and identify current research gaps that still hinder the development of practical \ac{SMC}-based applications.
Our survey consists of two parts; this paper and a companion paper.
While the companion paper focuses on \ac{SMC} with relatively long communication ranges, this paper covers \ac{SMC} over short distances of typically not more than a few millimeters.
\vspace*{-0.4cm}
\end{abstract}
\acresetall

\section{Introduction}
\label{sec:introduction}
The term {\em \ac{SMC}} labels an ongoing effort in the communications engineering community to conquer the realm of natural {\em chemical communication}.
Chemical communication denotes the exchange of information by means of chemical messengers, i.e., molecules, and it occurs in a huge variety of forms and across a wide range of scales in nature; examples include the pheromone-based communication between different individuals of the same animal species, the hormone signaling inside the human body, and the intercellular signaling in bacteria colonies \cite{nakano13}.
We refer to natural chemical communication systems in the following as {\em \ac{NMC}} systems.
Inspired by \ac{NMC}, the research field of \ac{SMC} revolves around the idea of establishing {\em synthetic} communication links by leveraging chemical means of communication \cite{nakano13}.
Specifically, and in contrast to conventional synthetic communication based on \ac{EM}, \ac{SMC} employs molecules to convey information.
In this sense, \ac{SMC} is a bio-inspired communication paradigm.

From an application perspective, the main motivation of \ac{SMC} is to enable synthetic communication for scenarios that hamper or even prohibit the use of \ac{EM}-based communication.
One important class of such applications arises in the context of the {\em \ac{IoBNT}}, a communication network that integrates in-body nanodevices, i.e., devices of sizes smaller than $\SI{1}{\micro\meter}$, that cooperatively perform sensing and actuation tasks inside the human body \cite{akyildiz15}.
In the context of the \ac{IoBNT}, the motivation for \ac{SMC} is two-fold.
First, \ac{EM}-based communication suffers from high signal attenuation in human tissue.
The signal attenuation is particularly severe at the high carrier frequencies required by nanoscale antennas and hampers the realization of \ac{EM}-based in-body communication.
Second, equipping nanodevices with \ac{SMC} capabilities opens up the space for applications that require these devices to interface with natural communication peers, such as natural cells or organs; examples for such applications include brain-machine interfaces \cite{Veletic2019} and targeted drug delivery \cite{Felicetti2016}.
Other classes of applications for \ac{SMC} arise in the context of communication in \ac{EM}-denied environments outside the human body such as in oil pipelines, where \ac{EM}-based communication is inefficient, or in industrial environments processing explosive gases, where \ac{EM}-based communication may not be allowed.
In light of this discussion, \ac{SMC} promises to extend the reach of synthetic communication beyond the boundaries of today's prevalent synthetic communication paradigms.

In the quest to realize the envisioned revolutionary \ac{SMC}-based applications, researchers have tackled the field from different angles using a variety of different research methods.
One line of research has focused on the development of theoretical models to explore the fundamental limits of \ac{SMC} and devise design guidelines for \ac{SMC} devices and systems; these efforts have been surveyed comprehensively, for example, \mbox{in \cite{jamali19, kuscu19}}.

At the same time, it was understood early on that {\em experimental research} was needed to complement the ongoing theoretical research efforts in \ac{SMC}.
However, experimental research to date is still rather underrepresented in \ac{SMC}, as evidenced by the small number of experimental contributions relative to the purely theoretical ones.
Now, on the one hand, the experimental domain is less accessible than the theoretical one to many \ac{SMC} researchers who mostly have a background in communications engineering, but limited experience, expertise, and facilities required to conduct experiments involving chemical or biological system components.
On the other hand, the development of practical \ac{SMC}-based applications in the future will require a tight integration of theoretical and experimental research efforts.
In fact, several studies in the \ac{SMC} literature already showcase the potential of the fruitful interaction between theoretical and experimental research \mbox{\cite{austin14,grebenstein18,terrell21}}.

This survey presents a step towards making experimental \ac{SMC} research more accessible and in this way foster its permanent integration into the mainstream of \ac{SMC} research.
To this end, 
we provide a systematic review of existing experimental \ac{SMC} research that allows us to contrast the current state of the art with the envisioned applications and identify major existing research gaps.
The survey is organized in two parts.
Part~I, this paper, presents an overview of general concepts in \ac{SMC}, an exposition of the main \ac{SMC} application scenarios, and a comprehensive review of experimental \ac{SMC} works targeting applications with short communication distances, i.e., distances of not more than a few millimeters.
Part II presents a comprehensive review of experimental \ac{SMC} works related to long-range \ac{SMC} applications, i.e., applications requiring communication over distances of more than a few millimeters.
This second part of our survey is covered in a companion paper that appears alongside this paper in the same magazine and issue.

The remainder of this paper is organized as follows.
In the rest of this section, existing \ac{SMC} surveys related to experimental contributions are briefly reviewed and the outline of this survey is presented.
In Section \ref{sec:concepts_and_taxonomy}, a concise taxonomic framework including the most relevant physical processes and the main application domains for \ac{SMC} is introduced.
In Sections \ref{sec:short-range:in-body} and \ref{sec:short-range:loc}, experimental works on short-range \ac{SMC} related to in-body and lab-on-chip are discussed, respectively.
Finally, Section~\ref{sec:conclusions_short-range} concludes this paper with  an outlook on future research directions for short-range \ac{SMC}.

\subsection{Existing Surveys on Molecular Communications}

The existing survey papers, which have been published over the past years, span a variety of different topics and focus on different aspects of \ac{SMC}, e.g., future applications, networking aspects or physical layer channel modeling \cite{Farsad_et_al:IEEECommSurvTutorials:2016, Soldner2020,jamali19}. In this survey paper, we focus on experimental research in \ac{SMC}. Therefore, we will only discuss other survey papers that focus on experimental testbed design, interfacing between domains, and the design of practical components, in more detail (see Table~\ref{tab:surveys})\footnote{For a comprehensive overview of existing surveys and tutorials on \ac{SMC}, we refer the reader to the overview tables in \cite{Soldner2020,Bi2021}.}. 
In \cite{Darchini2013}, networking aspects are discussed and a particular focus is on communication via microtubules. Interfacing aspects are discussed in \cite{Veletic2019, Jornet2019, Kim2019}, where \cite{Veletic2019} focuses on synaptic communication and the design of brain-machine interfaces, in \cite{Jornet2019}, optical nano-bio interfaces are discussed for the connection of biological and electrical networks, and \cite{Kim2019} proposed methods for the interfacing with biological systems. The design of components for practical \ac{SMC} systems is discussed in \cite{Soldner2020, kuscu19}, where \cite{kuscu19} describes the physical design of practical \acp{Tx} and \acp{Rx} for \ac{SMC}, and \cite{Soldner2020} investigates possible biological building blocks for the design of SMC systems. Surveys \cite{Yang2020, Huang2021} also provide an overview of a subset of experimental works in \ac{SMC}. In particular, the authors in \cite{Yang2020} provide an overview on molecular testbeds and experiments and  \cite{Huang2021} gives an overview on macroscale \ac{SMC} testbeds which are categorized by the channel type. 

\begin{table*}[t]
\caption{\small Existing surveys focusing on experimental testbed design, interfacing, and the design of practical components.}
\vspace*{-0.2cm}
\label{tab:surveys}
\centering
\begin{tabular}{p{3.3cm}c p{13cm}}
\hline\noalign{\smallskip}
 \multicolumn{1}{c}{Reference}& \multicolumn{1}{c}{Year} & \multicolumn{1}{c}{Detailed Notes}\\[0.1em]
\noalign{\smallskip}\hline\noalign{\smallskip}
Darchini et al. \cite{Darchini2013} & 2013 & Focus on nanonetworks and practical communication via microtubules and physical contact. \\[0.5em]
Veletic et al. \cite{Veletic2019} & 2019 & Synaptic communication. Design of brain-machine interfaces.\\[0.5em]
Kim et al. \cite{Kim2019} & 2019 & Redox reactions as modality for propagating electrical and biological signals.\\[0.5em]
Kuscu et al. \cite{kuscu19} & 2019 & \mbox{}\Ac{Tx} and \ac{Rx} design for \ac{SMC} systems including modulation, coding, and detection schemes.
Focus on physical design of \ac{Tx} and \ac{Rx} and the concomitant challenges and limitations.
\\[0.5em]
Jornet et al. \cite{Jornet2019} & 2019 & Connection of biological networks to electronic computing systems by optical nano-bio interfaces.\\[0.5em]
S\"{o}ldner et al. \cite{Soldner2020} & 2020 & Biological components as building blocks (\ac{Tx}, \ac{Rx}, information molecules) for \ac{SMC} systems.\\[0.5em]
Yang et al. \cite{Yang2020} & 2020 & 
Testbed review. Recent developments of \ac{EM} and \ac{SMC}. 
Focus on applications and network structures. 
Hybrid communication paradigms. Inter-connectivity of EM and \ac{SMC} networks.\\[0.5em]
Huang et al. \cite{Huang2021} & 2021 & Macro-scale \ac{SMC} testbeds. Categorization based on channel type, i.e., gaseous and liquid channels.\\[0.5em]
Lotter et al. (this survey) & 2023 & Comprehensive overview of experimental works in \ac{SMC} and systematic classification into short-range (Part I) and long-range (Part II) \ac{SMC} systems.\\
\noalign{\smallskip}\hline\noalign{\smallskip}
\end{tabular}
\end{table*}

\subsection{Outline of This Survey}

\begin{figure*}
    \centering
    \includegraphics[width=\textwidth]{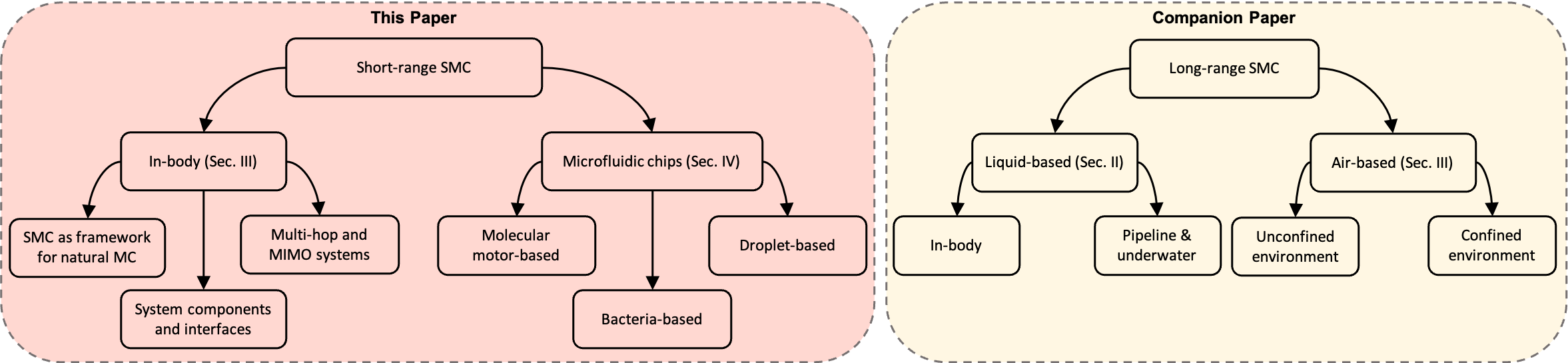}
    \vspace*{-0.6cm}
    \caption{Classification of experimental works and organization of the survey. This paper (Part I) reviews short-range \ac{SMC}, while the companion paper (Part II) targets long-range \ac{SMC}.
    }
    \label{fig:classification}
\end{figure*}

This two-part survey is organized according to the classification of the target applications for \ac{SMC} along different dimensions.
The applied classification scheme along with the corresponding sections of the survey is presented in Fig.~\ref{fig:classification}.

On the top level, we distinguish applications that require communication over distances of more than a few millimeters from those that do not.
We refer to the former ones (covered in the companion paper) as {\em long-range} and the latter ones (this paper) as {\em short-range}.
For short-range applications, we identified two application types, one focusing on communication inside the human body, i.e., {\em in-body} communication, cf.~Section~\ref{sec:short-range:in-body}, the other one on {\em microfluidic chips}, cf.~Section~\ref{sec:short-range:loc}.
These two categories are further subdivided according to the different scopes of the experimental contributions in each category.
For the long-range applications, we distinguish different applications based on the medium in which the communication takes place.
Long-range \ac{SMC} systems based on {\em liquid} media are reviewed in the companion paper in Section~II; those based on airborne communication, i.e., {\em air-based} \ac{SMC} systems, are reviewed in the companion paper in Section~III.
Furthermore, the works on liquid-based and air-based long-range \ac{SMC} reviewed in the companion paper are further categorized into different subcategories according to the respective propagation environment.

Applying the above classification scheme, this survey identifies the main drivers of experimental research in \ac{SMC} in terms of targeted applications in the past, current challenges, and opportunities for further progress in the field.

\section{Taxonomy: Physics and Applications}
\label{sec:concepts_and_taxonomy}

\begin{figure*}
    \centering
    \includegraphics[width=0.8\textwidth]{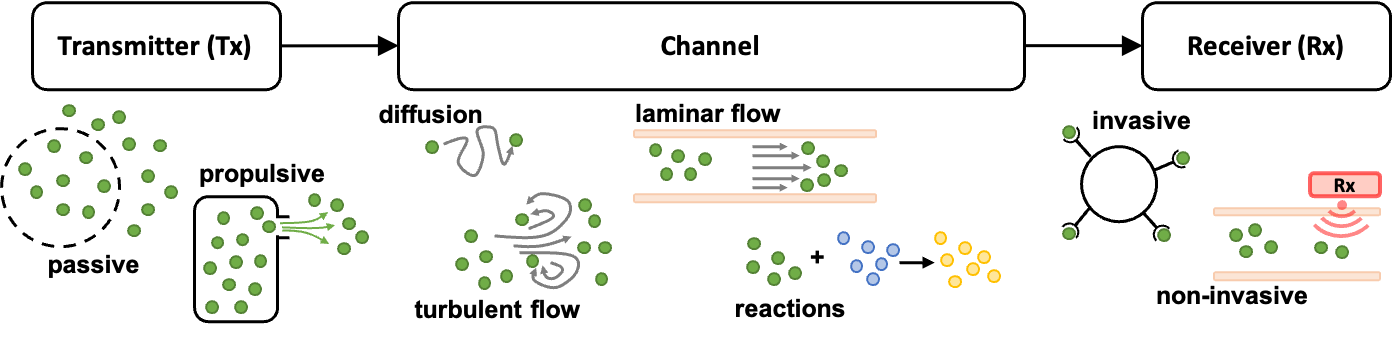}
    \vspace*{-0.4cm}
    \caption{Illustration of the main building blocks of an SMC system (namely, Tx, channel, and Rx) and the key physical principles governing the release, propagation, and reception processes of information molecules.}
    \label{fig:physics}
\end{figure*}

\subsection{Physical Governing Principles}
\label{sec:taxonomy:physics}

\Ac{SMC} systems concern the release,  propagation, and  reception of information-carrying molecules, which are referred to as {\em information molecules}\footnote{The term ``molecule'' is not used in a strict sense in this context but rather in a broad sense referring to any chemical substance employed as an information carrier. Examples include hydrogen ions (i.e., subatomic particles) \cite{grebenstein18}, nucleic acids (i.e., large biomolecules) \cite{furubayashi18}, and artificial nanoparticles \cite{nakano14}.}. In this process, the information may be embedded into 
(i) the presence and absence, or, more generally, the {\em concentration} or {\em relative concentrations} of the information molecule(s), 
(ii) the {\em type} of information molecule(s),
(iii) the {\em time} at which information molecules are released into the environment, or
(iv) the {\em location} from which information molecules are released \cite{jamali19}.
In most works reviewed in this survey, category (i) is assumed, where molecules are released (not released) by the \ac{Tx} if a binary '1' ('0') is to be transmitted.
This modulation scheme is referred to as {\em \ac{OOK}}.

Some relevant physical principles governing the release, propagation, and reception of information molecules in \ac{SMC} systems are illustrated in Fig.~\ref{fig:physics} and briefly discussed in the following:  

\textbf{Release mechanism:}
The \ac{Tx} releases the information molecules into the channel in a controlled manner according to the adopted modulation scheme. Examples of release mechanisms include: 
\begin{itemize}
    \item {\em passive} release where gates are opened at the \ac{Tx} and information molecules passively diffuse into the channel (e.g., via voltage-gated ion channels), and
    \item {\em propulsive} release where the information molecules are pushed into the channel (e.g., via a pump or spray).  
\end{itemize}

\textbf{Propagation channel:}
Depending on the application of interest, the propagation medium may be liquid or air. The propagation behavior of the information molecules within the fluid channel is affected by various physical phenomena including \cite{jamali19}: 
\begin{itemize}
    \item {\em free diffusion} due to thermal vibrations and collisions with
other molecules,
    \item {\em active mass transport} by either a force-induced drift or a bulk background flow, or
    \item {\em chemical reactions} that may degrade the information molecules or transform them into another type of molecules. 
\end{itemize}
Background flow plays an important role for medium- and long-range \ac{SMC} and can be further categorized into turbulent and laminar depending on whether  the variations in the flow velocity, over space and/or time, are stochastic (e.g., due to rough surfaces), or deterministic, respectively.  Turbulent flow is more prevalent in large-scale liquid-based and air-based \ac{SMC} systems whereas laminar flow is more common in  long-range pipe-shape \ac{SMC} channels (e.g., most blood vessels). 

\textbf{Reception mechanism:} The \ac{Rx} detects the presence of information molecules within its vicinity, which can be achieved via either  
\begin{itemize}
    \item  {\em invasive} reception where (part of) the \ac{Rx} device is located inside the communication medium to directly detect the information molecules (e.g., via chemical reactions between the molecules and the \ac{Rx} \cite{terrell21}), or
    \item  {\em non-invasive} reception where information molecules are indirectly detected by an external \ac{Rx} device (e.g., external observation of fluorescence emission by information molecules \cite{kuscu15}).  
\end{itemize}

\subsection{Main SMC Application Scenarios}
\label{sec:taxonomy:applications}

Motivated by the classification scheme introduced in Section~\ref{sec:introduction}, we provide here a brief review of the main application scenarios for \ac{SMC}, see also Fig.~\ref{fig:classification}.

\textbf{In-body SMC:}
One of the main motivations for \ac{SMC} are in-body applications, mostly inspired by the concept of the \ac{IoBNT} \cite{akyildiz15}.
In these applications, \ac{SMC} is employed to convey information between in-body devices such as mobile nanosensors that patrol the human blood stream or nanoparticles that cooperatively deliver medical agents.
In-body applications require small form factors of the communicating \ac{SMC} devices.
In addition, the devices need to be biocompatible \cite{Yang2020}.
There has been a wide range of devices suggested for the implementation of biocompatible nanodevices with communication capabilities.
Broadly, the proposed devices can be categorized into biological devices and artificial devices.
Biological devices encompass mainly synthetic cells.
Non-biological devices include polymers, hydrogels, and other synthetically produced materials.

\textbf{Lab-on-a-chip:}
\Ac{LoC} refers to the concept of integrating lab processes, such as sample analysis or preparation, on a microfluidic chip, i.e., a chip that comprises micro-scale components for handling tiny amounts of liquid, such as channels, mixers, valves, and pumps \cite{kirby10}.
Advantages of \acp{LoC} as compared to conventional labs are typically reduced costs and small form factors.
These enable, for example, novel diagnostic tools that can be applied in areas where large medical lab facilities are not available.

\textbf{Pipeline and underwater communications:}
Some applications for \ac{SMC} in liquid media are expected on very large scales, such as in pipelines or in the open sea.
One motivation of these applications is to exploit the reduced propagation loss of \ac{SMC} as compared to \ac{EM}-based or acoustic communication links for underwater search-and-rescue applications.
Another motivation is to provide covert communication links or enable search-and-rescue missions where the search target emits a molecular signal that is sensed by a rescue agent.

\textbf{Air-based communication:}
\ac{SMC} in gaseous environments, i.e., air-based \ac{SMC}, is envisioned to enable long-range wireless communication for scenarios in which \ac{EM}-based communication is not available.
The operation of a robot swarm in an emergency area, for example, could be coordinated via \ac{SMC}, where one robot upon reaching a missing person guides the other robots by releasing information molecules.
The inspiration of most air-based \ac{SMC} applications is drawn from nature, where the behavior of individuals within a group is coordinated via pheromone communication.

\section{Short-Range In-Body SMC}
\label{sec:short-range:in-body}

In this section, we review experimental works related to short-range in-body \ac{SMC}.
In this context, experiments have been employed to (i) understand and develop models for short-range in-body \ac{NMC} systems, (ii) develop and test the design of individual system components for \ac{SMC}, and (iii) study \ac{SMC} systems that involve multiple single communication links.

\subsection{SMC as Modeling Framework for NMC}

Some works in the context of short-range \ac{SMC} reproduce \ac{NMC} processes in experiments to aid the development of communication theoretic models for these processes.
Since \ac{NMC} processes commonly depend on many different unknown parameters, physics-based bottom-up modeling of these processes is often impossible.
Accurate models could, however, help to devise novel diagnosis and treatment strategies for diseases related to \ac{NMC} processes.
To overcome this problem, the works presented in this section explore \ac{SMC} as a data-aided modeling framework for \ac{NMC} processes.

\textbf{Cardiovascular disease:} The role of \ac{NMC} in the early stages of atherosclerosis, a common cardiovascular disease, was studied in \cite{felicetti14}.
Specifically, the \ac{NMC} between platelets and endothelial cells in blood vessels was reproduced experimentally; it was shown that the presence of a particular molecular signal that is emitted by platelets upon receiving an external stimulus leads to the over-expression of cell-adhesion molecules by endothelial cells, possibly promoting atherosclerosis.

\textbf{Brain cancer:} In \cite{awan21}, an \ac{SMC} model for the growth of glioblastoma multiforme, an aggressive form of cancer that originates in the brain, was developed. 
In particular, the structural growth pattern of glioblastoma multiforme cells observed in cell culture guided the theoretical \ac{SMC} modeling.

The two papers reviewed above showcase that experimental research can indeed drive the communication theoretic modeling of \ac{NMC} systems in the framework of \ac{SMC}.
It remains to be shown, however, that such modeling efforts will eventually aid the development of efficient diagnosis and treatment strategies for \ac{NMC}-related diseases.
In summary, the full potential of experiment-driven \ac{SMC} research for the characterization of \ac{NMC} processes (and eventually their therapeutic manipulation) has not been fully explored, yet.

\subsection{Design of System Components and Interfaces}

\begin{figure*}
    \centering
    \includegraphics[width=\textwidth]{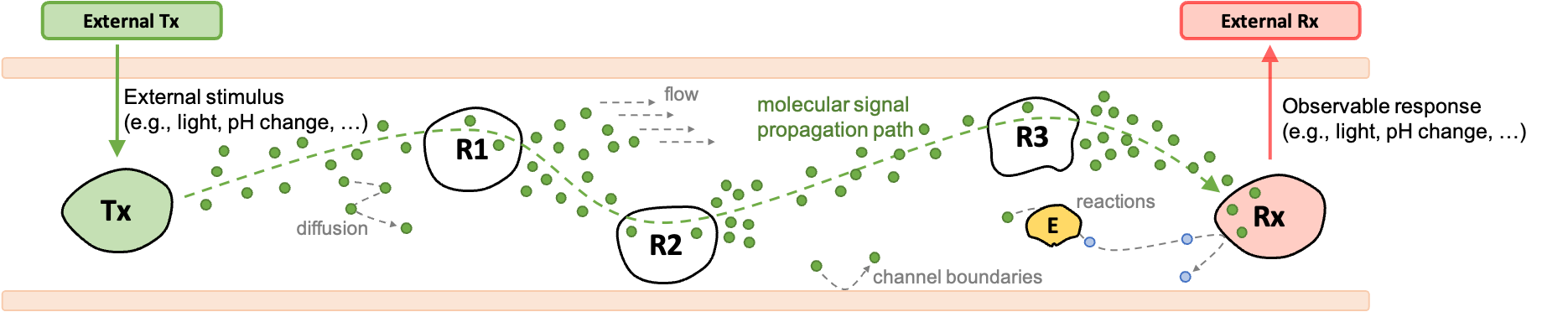}
    \vspace*{-0.6cm}
    \caption{Short-range in-body \ac{SMC} shown schematically. In response to an external stimulus, a \ac{Tx} entity generates a molecular transmit signal that propagates to a \ac{Rx}. Along the way from the \ac{Tx} to the \ac{Rx} the signal is possibly amplified by one or more relays (R1 - R3) and affected by effects such as diffusion, flow, and chemical reactions such as enzyme (E) reactions. The \ac{Rx}, finally, responds to the molecular signal by generating an externally observable response.}
    \label{fig:short-range_in-body}
\end{figure*}

In short-range in-body \ac{SMC}, a nanoscale \ac{Tx} generates a molecular signal in response to an external stimulus (control signal).
The molecular signal propagates in the environment where it is possibly amplified by one or more relays.
Finally, when the signal reaches the \ac{Rx}, it triggers an externally observable output signal or some other response.
This process is depicted in Fig.~\ref{fig:short-range_in-body}.

Different physical designs have been explored for the implementation of short-range in-body \ac{SMC} systems in the literature.
The different designs range from fully {\em artificial devices} to {\em biological cells}.
Artificial device designs comprise non-living objects such as single molecules or functionalized nanoparticles.
Biological cell-based designs utilize living cells, mostly bacteria cells, as \ac{SMC} \acp{Tx} and \acp{Rx}; the biological cells used in these designs are usually derived from natural cells by appropriate genetic modification.
{\em Hybrid} artificial/biological approaches that integrate artificial and biological components into one single device are also possible.

Specific challenges for the design of short-range in-body \ac{SMC} system components include (i) how to generate the molecular signal (transmit signal) at the \ac{Tx} (\ac{Tx} design), (ii) what type of transmit signal to generate (transmit signal design), and (iii) how to observe and process the molecular signal at the \ac{Rx} (\ac{Rx} design).
The review of related works in this section is organized according to these challenges.

\subsubsection{Tx Design} Here, we review different \ac{Tx} designs in the literature employing hydrogel-based devices and bacteria, respectively. 

\textbf{Hydrogel-based \acp{Tx}:} %
The artificial and hybrid \ac{Tx} designs proposed in \cite{liu15} are based on {\em hydrogel beads}.
Hydrogel beads are spherical containers that can, for example, be fabricated from alginate-based hydrogels. %
In \cite{liu15}, hydrogel beads were functionalized in two different ways to act as artificial and hybrid \ac{SMC} \acp{Tx}, respectively.
For the artificial \ac{Tx}, hydrogel beads were embedded with surface-proteins that catalyze the reaction of environment molecules (external stimulus) to information molecules.
For the hybrid \ac{Tx}, biological cells producing information molecules were directly embedded into the hydrogel beads.
The successful signal transmission was confirmed experimentally for both proposed \ac{Tx} designs in \cite{liu15}.

A hybrid \ac{Tx} based on hydrogel was presented in \cite{martins20}.
In \cite{martins20}, \ac{sfGFP} producing biological cells (bacteria) were embedded in hydrogel.
Hereby, the production of the primary \ac{sfGFP} signal was facilitated by a secondary, \ac{NMC} process among the bacteria.
Furthermore, it was suggested in \cite{martins20} that the hydrogel environment aided this secondary communication process, leading to improved \ac{SMC} signal generation.
Wet lab experiments confirmed the feasibility of the proposed \ac{Tx} design in \cite{martins20}.

\textbf{Bacteria-based \ac{Tx}:} In \cite{grebenstein18,grebenstein19}, genetic engineering was utilized to implement a biological \ac{SMC} \ac{Tx}.
To this end, bacteria were embedded with light-driven proton pumps, i.e., proteins that, upon receiving a light stimulus, pump protons from the interior of the bacteria into the environment.
The change in the pH value around the bacteria resulting from their proton pumping activity was sensed by a pH meter.
Finally, the biological \ac{Tx} in \cite{grebenstein18,grebenstein19} was used to transmit binary data using \ac{OOK} and this data was successfully detected from the received pH signal.

\subsubsection{Rx Design}
In the following, we review Rx designs based on polystyrene, hydrogel, and engineered bacteria, respectively.%

\textbf{Polystyrene-based \ac{Rx}:} An artificial \ac{SMC} \ac{Rx} design was proposed in \cite{nakano14}.
In this design, polystyrene beads, i.e., spheres assembled from non-biological material, were functionalized with fluorescent pH indicator proteins and introduced into biological cells.
It could be observed in wet lab experiments that an externally observable optical signal was indeed generated by the \ac{SMC} \ac{Rx} in response to an intracellular chemical signal.

\textbf{Hydrogel-based \ac{Rx}:} A hybrid \ac{SMC} \ac{Rx} was presented in \cite{liu15}.
Here, hydrogel beads were embedded with fluorescent reporter cells and it was demonstrated that the proposed \ac{Rx} could successfully detect information molecules in its envrionment.

\textbf{Bacteria-based \acp{Rx}:} Two biological \ac{SMC} designs were proposed in \cite{sezgen21}.
In both designs, bacteria were used to sense a molecular signal of interest, e.g., the presence of a biomarker, and the resulting bacteria activity was then detected externally as a proxy for the molecular signal.
In the first design proposed in \cite{sezgen21}, engineered bacteria were co-located with an LED and a photodiode on an in-body implant capable of actively transmitting wireless \ac{EM} signals.
In this design, the bacteria express fluorescent proteins in response to a molecular signal.
The proteins are then illuminated by the LED at a specific optical wavelength and the resulting fluorescence signal (at a different wavelength) is converted to an electrical signal by the photodiode.
The electrical signal is finally transmitted to an external device using wireless \ac{EM}-based communication.
The second \ac{Rx} design proposed in \cite{sezgen21} relies on backscatter communication.
Here, bacteria are co-located with a degradable resonator on a passive implant.
The degradation of the resonator in this design is dependent on the bacteria activity.

\subsubsection{Signal Design} Finally, two works consider the \ac{SMC} transmit signal design \cite{martins18,martins21}.
Specifically, the challenges of (i) choosing the appropriate type of \ac{SMC} information molecules and (ii) redundantly encoding information in an \ac{SMC} \ac{Tx} were considered in \cite{martins18} and \cite{martins21}, respectively.

\textbf{SMC for preventing biofilm formation by bacteria:} Which type of information molecules to use in a particular \ac{SMC} system can depend on the specific target application.
In \cite{martins18}, \ac{SMC} is utilized as a tool to prevent biofilm formation by bacteria.
Specifically, bacteria rely on \ac{NMC} to facilitate biofilm formation and the \ac{SMC} system in \cite{martins18} is designed to disturb this \ac{NMC} system.
To this end, wet lab experiments were conducted in \cite{martins18} for development of an \ac{SMC} signal that could effectively accomplish this task.

\textbf{Repetition coding using multiple bacteria populations:} In \cite{martins21}, bacteria populations of two different types were employed to implement an \ac{SMC} channel coding scheme.
The idea of the proposed scheme is to redundantly encode one single input signal using two independent encoding devices (corresponding to the two bacteria populations).
In the experiments conducted in \cite{martins21}, one bacteria population produced a red fluorescence signal and the other one a green fluorescence signal.
The measured output signal was a combination of the two.
However, the two different bacteria types were induced with two different input signals and, hence, an end-to-end validation of the proposed coding scheme is still missing.

In summary, several different physical \ac{SMC} system component designs have been studied experimentally for short-range in-body \ac{SMC}.
However, the works presented in this section comprise mainly proof-of-concept experiments and feasibility studies.
It thus remains an open question, which component design will be the most favorable for specific in-body \ac{SMC} applications.

\subsection{Multi-Hop and MIMO Systems}

\begin{figure*}
    \centering
    \includegraphics[width=0.8\textwidth]{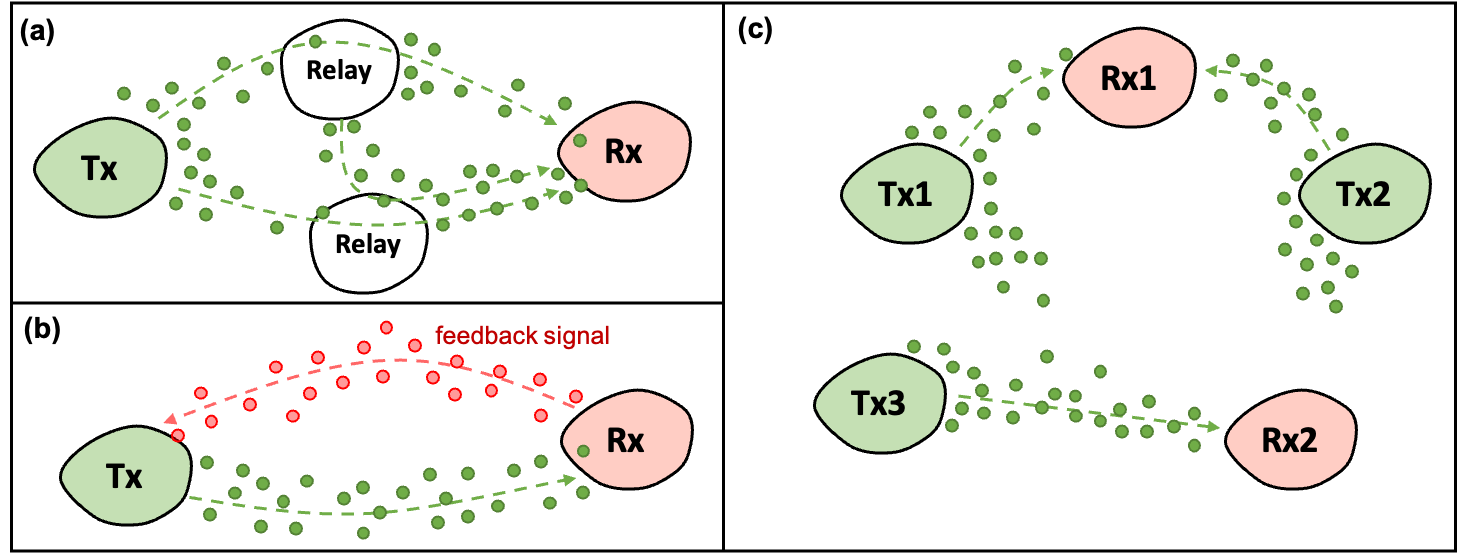}
    \vspace*{-0.3cm}
    \caption{Different system architectures for short-range in-body \ac{SMC} comprising multiple communication links. {\em (a):} The transmit signal is relayed by one (or more) {\em relay} nodes. {\em (b):} In response to receiving a signal from the \ac{Tx}, the \ac{Rx} sends a {\em feedback signal} to the \ac{Tx}. {\em (c):} Multiple nanodevices communicate the same information simultaneously in {\em \ac{MIMO}} \ac{SMC}.}
    \label{fig:srib_smc:multi-hop}
\end{figure*}

In this section, we review experimental works on short-range in-body \ac{SMC} in which the considered \ac{SMC} systems comprise more than one single physical communication link.
This includes multi-hop and \ac{MIMO} \ac{SMC} systems.
In multi-hop \ac{SMC}, multiple short-range \ac{SMC} links are cascaded to either achieve a larger communication range as compared to one-hop \ac{SMC} \cite{nakano07,nakano08,abbasi18,furubayashi18} or to provide a molecular feedback signal \cite{furubayashi18,terrell21}.
In \ac{MIMO} \ac{SMC}, multiple sender nanodevices (\ac{Tx} antennas) transmit the same data sequence simultaneously to multiple receiving nanodevices (\ac{Rx} antennas) \cite{kuscu15}.
Fig.~\ref{fig:srib_smc:multi-hop} illustrates these concepts schematically.

For the physical design of the \ac{Tx} and \ac{Rx} devices in the systems reviewed in this section, similar considerations as in the previous section apply.
In particular, both artificial and biological cell-based designs are considered in the reviewed works.

\subsubsection{Multi-hop \ac{SMC}} The following studies investigate multi-hop \ac{SMC} using water droplets and intra- and inter-cell communications, respectively.

\textbf{Water droplet-based RNA signaling:} %
Artificial devices were proposed in \cite{furubayashi18} for the implementation of a simple data link network protocol for multi-hop \ac{SMC}.
In the proposed network design, a \ac{RNA} message transmitted by a \ac{Tx} device is relayed by several transceiver devices in a decode-and-forward manner.
When the message finally reaches an \ac{Rx} device, a second type of \ac{RNA}, called ACK-RNA in \cite{furubayashi18}, is produced to confirm the successful end-to-end transmission.
Hence, in the proposed network protocol, the end-to-end communication range is increased by relaying {\em and} a feedback signal is provided.
The artificial transceiver devices in \cite{furubayashi18} are realized as water droplets that contain a minimal molecular engine able to translate and synthesize \ac{RNA}.
The wet lab experiments in \cite{furubayashi18} confirm the feasibility of the proposed relay and \ac{Rx} node designs, respectively.
End-to-end transmission in the proposed framework, however, is not confirmed experimentally in \cite{furubayashi18}.

\textbf{Multi-hop intra- and inter-cell communication:} %
Biological cell-based devices were utilized in the multi-hop \ac{SMC} designs in \cite{nakano07,nakano08,abbasi18,terrell21}.
In \cite{nakano07,nakano08}, human cancer cells expressing channel-proteins were arranged as a ``cell wire'' to relay an intracellular signal across multiple cells.
To this end, the channel-proteins form so-called {\em gap junctions}, i.e., tunnels that facilitate inter-cell communication.
Gap junction-based signaling is widely used by \ac{NMC} systems \cite{Bi2021}.
In \cite{nakano07,nakano08}, it was demonstrated that gap junctions could also be utilized to implement a multi-hop \ac{SMC} system.

In \cite{abbasi18}, the nervous system of a living roundworm was used to transmit information using neuro-spike communication.
In neuro-spike communication, molecular signals in the form of electro-chemical potentials are relayed along a row of neural cell segments \cite{Veletic2019}.
In each of the segments, the signal is amplified and forwarded to the next segment, such that reliable molecular signal transmission over a total range of up to more than one meter is possible.
In \cite{abbasi18}, two pin electrodes, a \ac{Tx} and an \ac{Rx} electrode, respectively, were inserted into a roundworm at its two opposite ends.
In this setup, the \ac{Tx} electrode was used to inject a \ac{FSK}-modulated binary signal into a specific part of the nervous system of the worm.
The average number of spikes was detected by the \ac{Rx} electrode as the received signal.
Based on this setup, comprehensive communication theoretic analysis was performed in \cite{abbasi18} and a peak data rate of $66.6\,\mathrm{bps}$ (bit per second) was achieved at a \ac{BER} of $6.8{\times}10^{-3}$.

In \cite{terrell21}, three different types of genetically engineered bacteria were utilized to implement a multi-hop \ac{SMC} system.
In the proposed setup, one bacteria colony (the \ac{Tx}) translates an electrical input signal by means of a redox reaction \cite{Kim2019} into a molecular transmit signal, i.e., it releases \ac{AHL} molecules.
This \ac{AHL} signal is then sensed by the two other bacteria colonies (the \acp{Rx}), a {\em verifier} colony, and an {\em actuator} colony.
The actuator colony is designed to release a therapeutic agent upon reception of the \ac{AHL} signal.
The verifier colony produces a molecular feedback signal that induces another redox reaction that is detected externally.
In this way, the connectivity of the proposed communication network is verified.

\subsubsection{MIMO \ac{SMC}} Next, we discuss \cite{kuscu15} that realizes a short-range \ac{MIMO} \ac{SMC} system.

\textbf{\ac{FRET}-based \ac{MIMO} \ac{SMC}:} A \ac{MIMO} \ac{SMC} system based on artificial \ac{Tx} and \ac{Rx} devices was proposed in \cite{kuscu15}.
Specifically, the \ac{SMC} \ac{Tx} and \ac{Rx} in \cite{kuscu15} consist of multiple fluorescent molecules.
The design of this system is based on the physical principle of {\em \ac{FRET}}.
\Ac{FRET} denotes the non-radiative energy transfer from one excited fluorescent molecule, the {\em donor}, to another, nearby fluorescent molecule in ground-state, the {\em acceptor}.
Immediately after such an energy transfer, the acceptor molecule is in excited-state and emits fluorescence, i.e., an optical signal, that can be detected by an external device.
In the communication system presented in \cite{kuscu15}, digital data was modulated onto an optical signal using \ac{OOK}.
The modulated optical signal was then passed on to the donor molecules.
Since the donor molecules were co-located with acceptor molecules in a bulk solution, donor molecules excited by the applied optical signal randomly transferred part of their energy to acceptor molecules using the \ac{FRET} mechanism.
In this way, a \ac{MIMO} \ac{SMC} link at nanoscale was realized.
Finally, the fluorescence signal captured from the acceptor molecules was used for signal detection in \cite{kuscu15}.
Using the setup just described, data rates of $50$, $150$, and $250\,\mathrm{kbps}$  could be achieved at \acp{BER} of $1.9 {\times} 10^{-6}$, $5.7 {\times} 10^{-5}$, and $3.1 {\times} 10^{-2}$, respectively.

In summary, it was demonstrated in the works reviewed in this section that the concepts of multi-hop and \ac{MIMO} \ac{SMC} could potentially be useful for future in-body applications based on short-range \ac{SMC}.
In particular it was shown that relaying, feedback, and the operation of multiple nanodevices in \ac{MIMO} mode could be used to enhance the communication range, the operation, and the data rate, respectively, of short-range \ac{SMC} systems.

\subsection{Summary}

Short-range in-body \ac{SMC} has been experimentally explored in the past to (i) develop models for \ac{NMC} systems, (ii) provide proof-of-concept implementations of single \ac{SMC} system components and interfaces, and (iii) develop prototypes for multi-hop and \ac{MIMO} \ac{SMC} systems, respectively.
The works reviewed in this section exemplify that different physical system designs for \ac{SMC} in each of these categories may be feasible.
However, which of the proposed designs is eventually most suitable depends on the requirements of the specific target applications.
In fact, a detailed specification of application-imposed constraints for the physical design of short-range in-body \ac{SMC} is still lacking to date; this makes it difficult to assess the potential of the different designs proposed in the literature for real-world applications.

\section{Lab-on-a-Chip Systems}
\label{sec:short-range:loc}

In this section, we review experimental works on \ac{SMC} targeting \ac{LoC} applications.
In the context of \ac{SMC}, \ac{LoC} applications have attracted research interest mainly in the following three different directions.

\textbf{Molecular motor-based \ac{LoC}:} In molecular motor-based \ac{LoC} systems, an analyte sample is transported by a molecular motor-based cargo transporter to one or more analysis sites.
Molecular motor-based transport is inspired by the intracellular transport of macromolecules and organelles by transport vesicles that are carried by molecules, e.g., kinesines.
The carrier molecules in these \ac{NMC} systems move along a {\em molecular rail}, e.g., a microtubule.
In one example application of molecular motor-based \ac{LoC}, the analyte sample is human sweat that is applied on a smartphone-embedded \ac{LoC} where an assay of specific biomarkers is conducted and the result is forwarded to a medical doctor. %

\textbf{Bacteria-based \ac{LoC}:} In bacteria-based \ac{LoC} systems, genetically engineered bacteria are used as biosensing devices to detect the presence of pollutants in environment monitoring assays. %
In order to process a particular analyte on a bacteria-based \ac{LoC}, specifically designed bacteria are embedded into a microfluidic chip and the analyte is delivered to them via the medium that flows through the chip.

\textbf{Droplet-based \ac{LoC}:} Finally, droplet-based \ac{LoC} systems constitute a class of microfluidic chips that allow for the injection of a {\em dispersed phase} into the background fluid (medium) of the microfluidic chip.
In these systems, the medium is called {\em continuous phase}, and the fluids of the dispersed and the continuous phase are immiscible.
The rationale behind droplet-based \ac{LoC} systems is to enable the integration of multiple modules onto one microfluidic chip in a programmable fashion.
This integration, in turn, requires the single modules to communicate with each other; to realize this communication, different ways of embedding information into droplets have been studied using \ac{SMC}.

\begin{figure*}
    \centering
    \includegraphics[width=\textwidth]{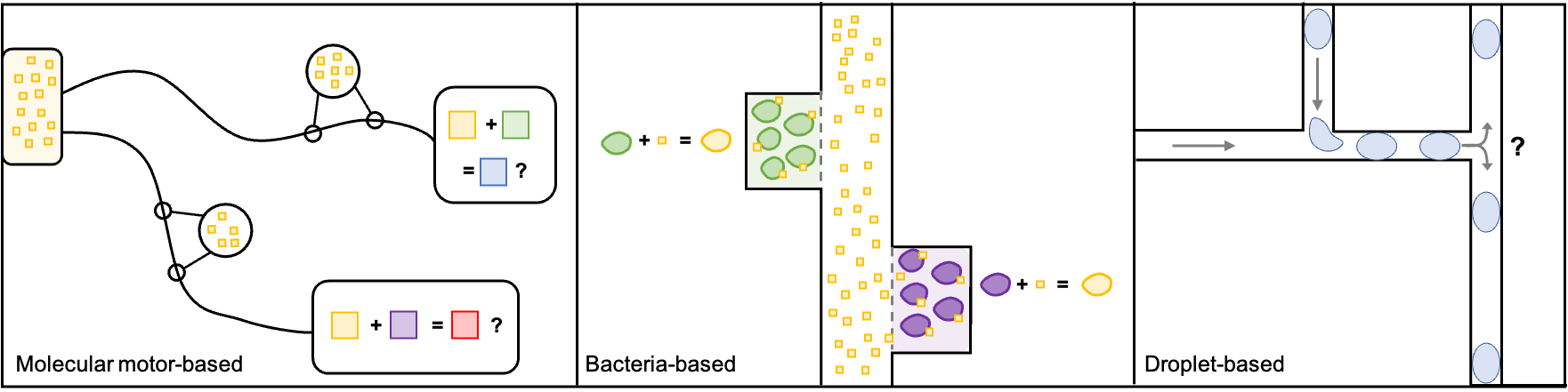}
    \vspace*{-0.5cm}
    \caption{Different \ac{SMC} applications for Lab-on-Chip systems: {\em Left:} Molecular motor-based biosensor assay. Analytes (orange) are carried towards assay sites by molecular motors. {\em Middle:} Bacteria communication-based LoC systems. Analytes are processed by bacteria housed in assay chambers. {\em Right:} Droplet-based microfluidics. Droplets (grey) are routed by hydrodynamics.}
    \label{fig:mfc-smc}
\vspace*{-0.3cm}
\end{figure*}

Fig.~\ref{fig:mfc-smc} illustrates the different classes of \ac{LoC} systems.

\subsection{Molecular Motor-based \ac{LoC}}

In this part, we discuss a line of early experimental works in \ac{SMC} \cite{sasaki06,moritani07,hiyama08,hiyama08a} that studied the use of tagged molecule containers (artificial vesicles) as transport vesicles for molecular motor-based \ac{LoC} applications.

\textbf{Controlled loading and delivery of liposome containers:} In a first feasibility study \cite{sasaki06}, it was observed that molecular tags could be used to enforce the delivery of tagged liposome containers to \ac{Rx} vesicles with matching labels.
In a further study \cite{moritani07}, the controlled loading of liposome containers with information molecules at a \ac{Tx} vesicle was achieved.
To this end, the transport vesicles as well as the \ac{Tx} vesicle were embedded with channel proteins that formed loading tunnels upon physical contact of the \ac{Tx} with the transporter vesicles.

\textbf{DNA-tagged molecular cargo transport:} The authors of \cite{hiyama08,hiyama08a} studied molecular motor-based active transport as a means to transport a molecular cargo from a designated molecular \ac{Tx} to a designated molecular \ac{Rx}.
Specifically, in the experimental setup in \cite{hiyama08,hiyama08a}, kinesines were immobilized onto a surface and microtubules were employed as mobile molecular cargo transporters.
Furthermore, both the molecular cargoes and the microtubules in \cite{hiyama08,hiyama08a} were labeled with \ac{ssDNA} and the selective loading of those cargoes labelled with \ac{ssDNA} complimentary to the microtubule label could be confirmed.
Also, the unloading of cargoes at specific \ac{ssDNA}-labeled \ac{Rx} sites could be demonstrated experimentally in \cite{hiyama08,hiyama08a}.
The loading and unloading processes hereby take place autonomously, i.e., they do not require external intervention.

It should be noted that, to the best of the authors' knowledge, the feasibility of the system designs studied in the works reviewed in this section was never verified in end-to-end experiments.
\vspace*{-0.3cm}

\subsection{Bacteria-based LoC}

The authors of \cite{krishnaswamy13,austin14,bicen15,austin17,martins22} employ bacteria as information \acp{Tx} or \acp{Rx} in an LoC setup.

\textbf{C6-HSL-sensitive bacteria:} In \cite{krishnaswamy13}, the transmission of information to bacteria cultured on a microfluidic chip was studied.
To this end, the bacteria were immobilized in trapping chambers on the microfluidic chip where they received an information-carrying molecular signal comprised of \ac{C6-HSL} molecules.
The signal was carried towards the bacteria by the fluid flow in the device.
Moreover, since the bacteria had been engineered to emit fluorescence upon stimulation with \ac{C6-HSL}, a fluorescence signal could be detected in \cite{krishnaswamy13} as output signal after the bacteria had received the molecular signal.
Using \ac{OOK} to modulate binary information onto the injection of \ac{C6-HSL} into the microfluidic device, the authors in \cite{krishnaswamy13} could demonstrate successful data transmission in the proposed setup.
However, the achieved data rate was only $3.8 {\times} 10^{-5}\,\mathrm{bps}$.
Motivated by this observation, the authors developed a modulation scheme to encode information into the time that elapses between two subsequent injections of information molecules.
The practical feasibility of this modulation scheme was, however, not demonstrated experimentally.

\textbf{AHL-sensitive bacteria:} In two further works \cite{austin14,bicen15} on bacteria-based \ac{LoC}, the response of bacteria to stimulation with a particular molecular signal was studied.
Specifically, the bacteria were engineered to express \acp{GFP} upon stimulation with \ac{AHL} molecules, one of the information molecules used in \ac{NMC} among bacteria. %
To this end, the \ac{NMC} apparatus of the bacteria was genetically modified.

In \cite{austin14}, a mathematical model of the intracellular signaling cascade leading to the expression of \acp{GFP} by the engineered bacteria was proposed and validated with experimental data.
The experimental data was collected from a microfluidic chip, similar to the one used in \cite{krishnaswamy13}, on which the bacteria had been incubated.
In a follow-up work \cite{bicen15}, different sampling strategies for the \ac{SMC} system considered in \cite{austin14} were developed and evaluated.
To this end, experimental data was used to assess the statistical properties of the bacteria response.

\textbf{AHL-based communication between bacteria colonies:} In \cite{austin17}, the communication between two bacteria colonies on a microfluidic chip in response to an external input signal was studied.
To this end, a \ac{Tx} and an \ac{Rx} bacteria colony, respectively, were grown in two distinct trapping chambers on the chip.
The two trapping chambers had been separated by a porous material.
The \ac{Tx} bacteria had been engineered to release an \ac{AHL} signal in response to stimulation with an externally induced molecular signal.
The bacteria in the \ac{Rx} colony in turn had been engineered to emit fluorescence upon exposure to an \ac{AHL} signal.
Upon stimulation of the \ac{Tx} colony, the \ac{AHL} molecules produced by the \ac{Tx} colony propagated through the porous barrier between the \ac{Tx} and the \ac{Rx} chambers via diffusion.
In this setting, externally controlled bacteria-to-bacteria communication could be observed experimentally in \cite{austin17}.

\textbf{Bacteria-based molecular computing:} Finally, a recent work studied the implementation of a biocomputing system using bacteria-based \ac{LoC} \ac{SMC} \cite{martins22}.
In this work, AND logic gates and ON-OFF switch sensors are realized using engineered bacteria.
Experimental results suggest that the secondary bacterial \ac{SMC} signal emitted in response to stimulation of the bacteria can be successfully detected by electrochemical sensors.

\subsection{Droplet-based \ac{LoC}}

The experimental implementation of routing protocol, modulation, and detection schemes for droplet-based \ac{LoC} have been investigated in \cite{deleo13,hamidovic19,bartunik20,bartunik20a}.

\textbf{Routing:} A routing protocol for networked droplet-based \ac{LoC} systems based on pure hydrodynamic control was proposed in \cite{deleo13}.
In this design, no external networking equipment is required to route droplets on a microfluidic chip.
Instead, the routing information is encoded into the relative distances between the single droplets.

\textbf{Modulation:} In \cite{hamidovic19}, different practical modulation schemes for the transmission of binary information in droplet-based \ac{LoC} systems were studied.
In particular, it was proposed to embed binary information into the presence/absence of droplets, the distance between droplets, or the droplet size.
The performance of the proposed modulation schemes in terms of the achievable information rate and the error tolerance was evaluated experimentally. %
In the experimental setup, the background flow is driven by an external pressure pump, the continuous phase and the droplets consist of oleic sunflower oil and deionized, colored water, respectively.
The droplet detection at the \ac{Rx} is carried out by a microscope and a high-speed camera.
It was observed that the distance-based modulation performed best in terms of both information rate and error tolerance, followed by the presence/absence-based modulation scheme, and the droplet size-based scheme.

\textbf{Enhanced detectors:} Experimental works on droplet-based \ac{LoC} systems within the past few years focus on improving droplet detection by using multiple sensors \cite{bartunik20} and machine learning-based detection methods \cite{bartunik20a}.

Besides the \ac{SMC}-based designs of droplet-based \ac{LoC} systems discussed in this section, a class of droplet-based \ac{LoC} systems based on so-called Belousov–Zhabotinsky oscillators has also been studied.
For a review of the motivation and the idea behind this systems, we refer the reader to \cite{torbensen17}.

\subsection{Summary}

Different approaches for \ac{SMC}-based \ac{LoC} applications have been studied over the past one and a half decades.
While the focus in the beginning was on molecular motor-based \ac{LoC} systems, this focus shifted towards bacteria-based and droplet-based \ac{LoC} systems.
Despite successful experimental feasibility studies, however, it remains to be shown that \ac{SMC} can indeed enable significant progress in these fields beyond the current state of the art.

\section{Conclusions and Outlook}
\label{sec:conclusions_short-range}
In this first part of our survey on experimental research in \ac{SMC}, we have reviewed experimental works related to short-range \ac{SMC}.
We have discussed different implementations of short-range \ac{SMC} systems for in-body applications.
In particular, we have seen that different types of \ac{Tx} and \ac{Rx} devices, ranging from artificial to biological ones, have been proposed and explored experimentally. Based on the reviewed works in this paper, the following directions for future research are identified:

\textbf{Optimized design of multi-node SMC:} Our review of experimental works on in-body \ac{SMC} in this paper suggests that practical in-body short-range \ac{SMC} will likely comprise several communication links to enhance the communication range, the achievable data rate, and the reliability of the system, respectively.
This finding suggests that communication theoretical models are required that embrace the possibility of multiple communication links.
In this way, the design of reliable and efficient multi-hop \ac{SMC} systems for practical applications will be further stimulated.

\textbf{Application-specific design and assessment of SMC systems:} The reviewed experimental studies are mainly feasibility studies; a systematic assessment of the advantages and disadvantages of particular \ac{SMC} system designs in the context of specific target applications is still missing.
For one class of in-body \ac{SMC} applications, for example, artificial devices may be suited best, while for another class of applications, hybrid or biological devices may be advantageous.
To address this issue, it would be instructive to identify specific benchmark applications for in-body \ac{SMC} and develop a catalog of application-specific requirements to assess the feasibility of experimental \ac{SMC} designs for these applications.
This also includes the definition of key performance indicators and metrics to evaluate and compare \ac{SMC} systems.

\textbf{Bacteria-based and droplet-based \ac{LoC}:} From this first part of our survey, we learned also that the field of bacteria-based and droplet-based \ac{LoC} applications remains largely unexplored in the context of \ac{SMC}.
In particular, only few different \ac{SMC} system designs have been studied for these systems and much is yet to be learned about the requirements of future real-world applications in this field.
To catalyze further progress in this direction, we envision that besides the ongoing experimental work, also further theoretical studies are required to reveal the fundamental design parameters and performance criteria of \ac{LoC} \ac{SMC} systems.

\textbf{Access to the experimental data on SMC:} Given the extensive efforts required for building an SMC testbed, the access to experimental platforms for SMC systems is expected to remain limited for many researchers in the MC community. One approach to alleviate this issue is to make the experimental data from the existing testbeds  available to the public. This helps to expand the impact of existing experimental research and promote future research on improving state-of-the-art experimental platforms. Examples of past initiatives in this direction include a competition based on shared experimental data in ACM NANOCOM 2021 (\texttt{http://nanocom.acm.org/nanocom2021/}).

\bibliographystyle{IEEEtran}    
\bibliography{IEEEabrv,references}

\begin{thebibliography}{10}
\providecommand{\url}[1]{#1}
\csname url@samestyle\endcsname
\providecommand{\newblock}{\relax}
\providecommand{\bibinfo}[2]{#2}
\providecommand{\BIBentrySTDinterwordspacing}{\spaceskip=0pt\relax}
\providecommand{\BIBentryALTinterwordstretchfactor}{4}
\providecommand{\BIBentryALTinterwordspacing}{\spaceskip=\fontdimen2\font plus
\BIBentryALTinterwordstretchfactor\fontdimen3\font minus
  \fontdimen4\font\relax}
\providecommand{\BIBforeignlanguage}[2]{{%
\expandafter\ifx\csname l@#1\endcsname\relax
\typeout{** WARNING: IEEEtran.bst: No hyphenation pattern has been}%
\typeout{** loaded for the language `#1'. Using the pattern for}%
\typeout{** the default language instead.}%
\else
\language=\csname l@#1\endcsname
\fi
#2}}
\providecommand{\BIBdecl}{\relax}
\BIBdecl

\bibitem{nakano13}
T.~Nakano, A.~W. Eckford, and T.~Haraguchi, \emph{Molecular
  Communication}.\hskip 1em plus 0.5em minus 0.4em\relax Cambridge University
  Press, 2013.

\bibitem{akyildiz15}
I.~F. {Akyildiz}, M.~{Pierobon}, S.~{Balasubramaniam}, and Y.~{Koucheryavy},
  ``The internet of bio-nano things,'' \emph{IEEE Commun. Mag.}, vol.~53,
  no.~3, pp. 32--40, Mar. 2015.

\bibitem{Veletic2019}
M.~Veletic and I.~Balasingham, ``Synaptic communication engineering for future
  cognitive brain-machine interfaces,'' \emph{Proc. IEEE}, vol. 107, no.~7, pp.
  1425--1441, 2019.

\bibitem{Felicetti2016}
L.~Felicetti, M.~Femminella, G.~Reali, and P.~Li{\`{o}}, ``{Applications of
  molecular communications to medicine: A survey},'' \emph{Nano Commun. Netw.},
  vol.~7, pp. 27--45, 2016.

\bibitem{jamali19}
V.~{Jamali}, A.~{Ahmadzadeh}, W.~{Wicke}, A.~{Noel}, and R.~{Schober},
  ``Channel modeling for diffusive molecular communication—{A} tutorial
  review,'' \emph{Proc. IEEE}, vol. 107, no.~7, pp. 1256--1301, Jul. 2019.

\bibitem{kuscu19}
M.~Kuscu, E.~Dinc, B.~A. Bilgin, H.~Ramezani, and O.~B. Akan, ``Transmitter and
  receiver architectures for molecular communications: A survey on physical
  design with modulation, coding, and detection techniques,'' \emph{Proc.
  IEEE}, vol. 107, no.~7, pp. 1302--1341, Jul. 2019.

\bibitem{austin14}
C.~M. Austin, W.~Stoy, P.~Su, M.~C. Harber, J.~P. Bardill, B.~K. Hammer, and
  C.~R. Forest, ``Modeling and validation of autoinducer-mediated bacterial
  gene expression in microfluidic environments,'' \emph{Biomicrofluidics},
  vol.~8, no.~3, p. 034116, 2014.

\bibitem{grebenstein18}
L.~Grebenstein, J.~Kirchner, R.~S. Peixoto, W.~Zimmermann, W.~Wicke,
  A.~Ahmadzadeh, V.~Jamali, G.~Fischer, R.~Weigel, A.~Burkovski, and
  R.~Schober, ``Biological optical-to-chemical signal conversion interface: A
  small-scale modulator for molecular communications,'' in \emph{Proc. 5th ACM
  Int. Conf. Nanosc. Comp. Commun.}, Reykjavik, Iceland, 2018.

\bibitem{terrell21}
J.~L. Terrell, T.~Tschirhart, J.~P. Jahnke, K.~Stephens, Y.~Liu, H.~Dong, M.~M.
  Hurley, M.~Pozo, R.~McKay, C.~Y. Tsao, H.-C. Wu, G.~Vora, G.~F. Payne, D.~N.
  Stratis-Cullum, and W.~E. Bentley, ``Bioelectronic control of a microbial
  community using surface-assembled electrogenetic cells to route signals,''
  \emph{Nat. Nanotechnol.}, vol.~16, no.~6, pp. 688--697, 2021.

\bibitem{Farsad_et_al:IEEECommSurvTutorials:2016}
N.~Farsad, H.~B. Yilmaz, A.~Eckford, C.~B. Chae, and W.~Guo, ``A comprehensive
  survey of recent advancements in molecular communication,'' \emph{IEEE
  Commun. Surv. Tuts.}, vol.~18, no.~3, pp. 1887--1919, 2016.

\bibitem{Soldner2020}
C.~A. S{\"{o}}ldner, E.~Socher, V.~Jamali, W.~Wicke, A.~Ahmadzadeh, H.-G.
  Breitinger, A.~Burkovski, K.~Castiglione, R.~Schober, and H.~Sticht, ``A
  survey of biological building blocks for synthetic molecular communication
  systems,'' \emph{IEEE Commun. Surv. Tuts.}, vol.~22, no.~4, pp. 2765--2800,
  2020.

\bibitem{Bi2021}
D.~Bi, A.~Almpanis, A.~Noel, Y.~Deng, and R.~Schober, ``{A survey of molecular
  communication in cell biology: Establishing a new hierarchy for
  interdisciplinary applications},'' \emph{IEEE Commun. Surv. Tuts.}, vol.~23,
  no.~3, pp. 1494--1545, 2021.

\bibitem{Darchini2013}
K.~Darchini and A.~S. Alfa, ``{Molecular communication via microtubules and
  physical contact in nanonetworks: A survey},'' \emph{Nano Commun. Netw.},
  vol.~4, no.~2, pp. 73--85, 2013.

\bibitem{Jornet2019}
J.~M. Jornet, Y.~Bae, C.~R. Handelmann, B.~Decker, A.~Balcerak, A.~Sangwan,
  P.~Miao, A.~Desai, L.~Feng, E.~K. Stachowiak, and M.~K. Stachowiak,
  ``Optogenomic interfaces: Bridging biological networks with the electronic
  digital world,'' \emph{Proc. IEEE}, vol. 107, no.~7, pp. 1387--1401, 2019.

\bibitem{Kim2019}
E.~Kim, J.~Li, M.~Kang, D.~L. Kelly, S.~Chen, A.~Napolitano, L.~Panzella,
  X.~Shi, K.~Yan, S.~Wu, J.~Shen, W.~E. Bentley, and G.~F. Payne, ``Redox is a
  global biodevice information processing modality,'' \emph{Proc. IEEE}, vol.
  107, no.~7, pp. 1402--1424, 2019.

\bibitem{Yang2020}
K.~Yang, D.~Bi, Y.~Deng, R.~Zhang, M.~M. {Ur Rahman}, N.~A. Ali, M.~A. Imran,
  J.~M. Jornet, Q.~H. Abbasi, and A.~Alomainy, ``{A comprehensive survey on
  hybrid communication in context of molecular communication and terahertz
  communication for body-centric nanonetworks},'' \emph{IEEE Trans. Mol. Biol.
  Multi-Scale Commun.}, vol.~6, no.~2, pp. 107--133, 2020.

\bibitem{Huang2021}
Y.~Huang, F.~Ji, M.~Wen, X.~Chen, Y.~Tan, and B.~Zheng, ``Survey on macro-scale
  molecular communication prototypes,'' \emph{SCIENTIA SINICA Informationis},
  vol.~51, no.~12, 2021.

\bibitem{furubayashi18}
T.~Furubayashi, Y.~Sakatani, T.~Nakano, A.~Eckford, and N.~Ichihashi, ``Design
  and wet-laboratory implementation of reliable end-to-end molecular
  communication,'' \emph{Wirel. Netw.}, vol.~24, no.~5, pp. 1809--1819, 2018.

\bibitem{nakano14}
T.~Nakano, S.~Kobayashi, T.~Suda, Y.~Okaie, Y.~Hiraoka, and T.~Haraguchi,
  ``Externally controllable molecular communication,'' \emph{IEEE J. Sel. Areas
  Commun.}, vol.~32, no.~12, pp. 2417--2431, 2014.

\bibitem{kuscu15}
M.~Kuscu, A.~Kiraz, and O.~B. Akan, ``Fluorescent molecules as transceiver
  nanoantennas: The first practical and high-rate information transfer over a
  nanoscale communication channel based on {FRET},'' \emph{Sci. Rep.}, vol.~5,
  no.~1, p. 7831, 2015.

\bibitem{kirby10}
B.~J. Kirby, \emph{Micro- and Nanoscale Fluid Mechanics: Transport in
  Microfluidic Devices}.\hskip 1em plus 0.5em minus 0.4em\relax Cambridge
  University Press, 2010.

\bibitem{felicetti14}
L.~Felicetti, M.~Femminella, G.~Reali, P.~Gresele, M.~Malvestiti, and J.~N.
  Daigle, ``Modeling {CD40}-based molecular communications in blood vessels,''
  \emph{IEEE Trans. NanoBiosci.}, vol.~13, no.~3, pp. 230--243, 2014.

\bibitem{awan21}
H.~Awan, A.~Odysseos, N.~Nicolaou, and S.~Balasubramaniam, ``Analysis of
  molecular communications on the growth structure of glioblastoma
  multiforme,'' in \emph{Proc. IEEE Global Commun. Conf.}, 2021, pp. 1--6.

\bibitem{liu15}
Y.~Liu, H.-C. Wu, M.~Chhuan, J.~L. Terrell, C.-Y. Tsao, W.~E. Bentley, and
  G.~F. Payne, ``Functionalizing soft matter for molecular communication,''
  \emph{ACS Biomater. Sci. Eng.}, vol.~1, no.~5, pp. 320--328, 2015.

\bibitem{martins20}
D.~P. Martins, H.~Q. O'Reilly, L.~Coffey, P.~D. Cotter, and S.~Balasubramaniam,
  ``Hydrogel-based bio-nanomachine transmitters for bacterial molecular
  communications,'' in \emph{Proc. 1st ACM Int. Workshop Nanosc. Comp. Commun.
  Appl.}, 2020, p. 14–19.

\bibitem{grebenstein19}
L.~Grebenstein, J.~Kirchner, W.~Wicke, A.~Ahmadzadeh, V.~Jamali, G.~Fischer,
  R.~Weigel, A.~Burkovski, and R.~Schober, ``A molecular communication testbed
  based on proton pumping bacteria: Methods and data,'' \emph{IEEE Trans. Mol.
  Biol. Multi-Scale Commun.}, vol.~5, no.~1, pp. 56--62, 2019.

\bibitem{sezgen21}
O.~F. Sezgen, O.~Altan, A.~Bilir, M.~G. Durmaz, N.~Haciosmanoglu, B.~Camli,
  Z.~C.~C. Ozdil, A.~E. Pusane, A.~D. Yalcinkaya, U.~O.~S. Seker, T.~Tugcu, and
  S.~Dumanli, ``A multiscale communications system based on engineered
  bacteria,'' \emph{IEEE Commun. Mag.}, vol.~59, no.~5, pp. 62--67, 2021.

\bibitem{martins18}
D.~P. Martins, K.~Leetanasaksakul, M.~T. Barros, A.~Thamchaipenet, W.~Donnelly,
  and S.~Balasubramaniam, ``Molecular communications pulse-based jamming model
  for bacterial biofilm suppression,'' \emph{IEEE Trans. NanoBiosci.}, vol.~17,
  no.~4, pp. 533--542, 2018.

\bibitem{martins21}
D.~P. Martins, J.~Drohan, S.~Foley, L.~Coffey, and S.~Balasubramaniam,
  ``Modulated molecular channel coding scheme for multi-bacterial
  transmitters,'' in \emph{Proc. 19th ACM Conf. Embed. Netw. Sens. Sys.},
  Coimbra, Portugal, 2021, p. 610–615.

\bibitem{nakano07}
T.~Nakano, T.~Suda, T.~Koujin, T.~Haraguchi, and Y.~Hiraoka, ``Molecular
  communication through gap junction channels: System design, experiments and
  modeling,'' in \emph{Proc. 2nd Bio-Insp. Models Netw. Inform. Comp. Sys.},
  Dec. 2007, pp. 139--146.

\bibitem{nakano08}
T.~Nakano, Y.-H. Hsu, W.~C. Tang, T.~Suda, D.~Lin, T.~Koujin, T.~Haraguchi, and
  Y.~Hiraoka, ``Microplatform for intercellular communication,'' in \emph{Proc.
  3rd IEEE Int. Conf. Nano/Micro Eng. Mol. Sys.}, 2008, pp. 476--479.

\bibitem{abbasi18}
N.~A. Abbasi, D.~Lafci, and O.~B. Akan, ``Controlled information transfer
  through an in vivo nervous system,'' \emph{Sci. Rep.}, vol.~8, no.~1, pp.
  1--12, Feb. 2018.

\bibitem{sasaki06}
Y.~Sasaki, M.~Hashizume, K.~Maruo, N.~Yamasaki, J.~Kikuchi, Y.~Moritani,
  S.~Hiyama, and T.~Suda, ``Controlled propagation in molecular communication
  using tagged liposome containers,'' in \emph{Proc. 1st Bio-Insp. Models Netw.
  Inform. Comp. Sys.}, 2006, pp. 1--1.

\bibitem{moritani07}
Y.~Moritani, S.~Hiyama, S.~Nomura, K.~Akiyoshi, and T.~Suda, ``A communication
  interface using vesicles embedded with channel forming proteins in molecular
  communication,'' in \emph{Proc. 2nd Bio-Insp. Models Netw. Inform. Comp.
  Sys.}, 2007, pp. 147--149.

\bibitem{hiyama08}
S.~Hiyama, T.~Inoue, T.~Shima, Y.~Moritani, T.~Suda, and K.~Sutoh, ``Autonomous
  loading, transport, and unloading of specified cargoes by using {DNA}
  hybridization and biological motor-based motility,'' \emph{Small}, vol.~4,
  no.~4, pp. 410--415, 2008.

\bibitem{hiyama08a}
S.~Hiyama, S.~Takeuchi, R.~Gojo, T.~Shima, and K.~Sutoh, ``Biomolecular
  motor-based cargo transporters with loading{/}unloading mechanisms on a
  micro-patterned {DNA} array,'' in \emph{Proc. IEEE Int. Conf. Micro Electro
  Mech. Syst.}, 2008, pp. 144--147.

\bibitem{krishnaswamy13}
B.~Krishnaswamy, C.~M. Austin, J.~P. Bardill, D.~Russakow, G.~L. Holst, B.~K.
  Hammer, C.~R. Forest, and R.~Sivakumar, ``Time-elapse communication:
  Bacterial communication on a microfluidic chip,'' \emph{IEEE Trans. Commun.},
  vol.~61, no.~12, pp. 5139--5151, 2013.

\bibitem{bicen15}
A.~O. Bicen, C.~M. Austin, I.~F. Akyildiz, and C.~R. Forest, ``Efficient
  sampling of bacterial signal transduction for detection of pulse-amplitude
  modulated molecular signals,'' \emph{IEEE Trans. Biomed. Circuits Syst.},
  vol.~9, no.~4, pp. 505--517, 2015.

\bibitem{austin17}
C.~M. Austin, D.~M. Caro, S.~Sankar, W.~F. Penniman, J.~E. Perdomo, L.~Hu,
  S.~Patel, X.~Gu, S.~Watve, B.~K. Hammer, and C.~R. Forest, ``Porous monolith
  microfluidics for bacterial cell-to-cell communication assays,''
  \emph{Biomicrofluidics}, vol.~11, no.~4, p. 044110, 2017.

\bibitem{martins22}
D.~P. Martins, M.~T. Barros, B.~J. O’Sullivan, I.~Seymour, A.~O’Riordan,
  L.~Coffey, J.~B. Sweeney, and S.~Balasubramaniam, ``Microfluidic-based
  bacterial molecular computing on a chip,'' \emph{IEEE Sens. J.}, vol.~22,
  no.~17, pp. 16\,772--16\,784, 2022.

\bibitem{deleo13}
E.~De~Leo, L.~Donvito, L.~Galluccio, A.~Lombardo, G.~Morabito, and L.~M.
  Zanoli, ``Communications and switching in microfluidic systems: Pure
  hydrodynamic control for networking labs-on-a-chip,'' \emph{IEEE Trans.
  Commun.}, vol.~61, no.~11, pp. 4663--4677, 2013.

\bibitem{hamidovic19}
M.~Hamidovi\'{c}, U.~Marta, G.~Fink, R.~Wille, A.~Springer, and W.~Haselmayr,
  ``Information encoding in droplet-based microfluidic systems: First practical
  study,'' in \emph{Proc. 6th ACM Int. Conf. Nanosc. Comp. Commun.}, Dublin,
  Ireland, 2019.

\bibitem{bartunik20}
M.~Bartunik, M.~Fleischer, W.~Haselmayr, and J.~Kirchner, ``Colour-specific
  microfluidic droplet detection for molecular communication,'' in \emph{Proc.
  7th ACM Int. Conf. Nanosc. Comp. Commun.}, 2020.

\bibitem{bartunik20a}
------, ``Advanced characterisation of a sensor system for droplet-based
  microfluidics,'' in \emph{Proc. IEEE SENSORS}, 2020, pp. 1--4.

\bibitem{torbensen17}
K.~Torbensen, F.~Rossi, S.~Ristori, and A.~Abou-Hassan, ``Chemical
  communication and dynamics of droplet emulsions in networks of
  {Belousov–Zhabotinsky} micro-oscillators produced by microfluidics,''
  \emph{Lab Chip}, vol.~17, pp. 1179--1189, 2017.

\end{thebibliography}

\end{document}